\documentclass[preprint]{aastex}

\usepackage{graphics}
\usepackage{tikz}
\usetikzlibrary{shapes,arrows}
\usetikzlibrary{shapes,snakes}
\usepackage{natbib}

\begin{document}

\title{Gamma ray attenuation in X-Ray binaries: An application to $LS\,I\,+61^\circ 303$}
\author{Paul D. Nu\~nez, Stephan LeBohec, Stephane Vincent}
\affil{Dept. of Physics \& Astronomy\\University of Utah\\115 S 1400 E, Salt Lake City, UT, USA.}

\begin{abstract}
The X-ray binary $LS\,I\,+61^\circ 303$, 
consisting of a main sequence Be star and a compact object has
been detected in the TeV range with MAGIC and VERITAS, and showed a clear
intensity modulation as a function of the orbital phase. We describe a
gamma-ray attenuation model and apply it to this system. Our first result is
that interaction of high energy photons with the background radiation produced
by the main sequence star alone does not account for the observed modulation. We then include
interactions between very high energy radiation and matter, and are able to
constrain fundamental parameters of the system such as the mass of the compact 
object and the density of circumstellar matter around the Be star. In our analysis of the TeV data,
we find that the compact object has mass $M_2>2.5M_{\odot}$ at the $99\%$ confidence level, 
implying it is most likely a black hole. However, we find a column density which conflicts with results from X-ray observations, suggesting that attenuation may not play an important role in the modulation.

\end{abstract}

\maketitle

\section{Introduction}

In the past few years, several high mass X-ray binaries have been detected as
gamma ray emitters \citep{Hess, Magic, Veritas}, causing an intensification of
observational and theoretical interest. High energy emitting binary 
systems consisting of a main sequence star and a compact object
are the only known  very high energy (VHE) galactic variable sources, and their short
periods of days or weeks make them even more interesting observational
targets. With increasing spectral coverage and statistics,
the nature of photon emission and absorption mechanisms is becoming increasingly 
constrained. Here we are concerned with high energy (TeV)
photons emitted from the vicinity of the compact object
and interacting with the background black body radiation and ejected material from the
companion star. Even though these systems can be incredibly complex, a simple model of the
absorption mechanisms and how they affect the system's 
light curve, can still shed light on many aspects such as the compact object mass
and the orbital parameters.\\

One such example is the high energy emitting binary $LS\,I\,+61^\circ 303$ \citep{Massi}.  It was first detected
in the TeV range with MAGIC \citep{Magic} and further observed with VERITAS
 at flux levels ranging between 5\% and 20\% of the Crab Nebula
\citep{Veritas}. This source has been observed throughout most of the
electromagnetic spectrum starting with radio frequencies and extending to VHE
 gamma rays \citep{Leahy}. This broad spectral study indicates
that the system consists of a main sequence Be star of mass $M_1=12.5\pm2.5\,M_{\odot}$ \citep{Casares}, surrounded by a circumstellar disk \citep{Grundstrom, Paredes}, and a compact companion
separated by tens of solar radii at periastron. The compact companion can be either a
neutron star or a black hole \citep{Casares}, and its exact nature 
is still subject of investigation and debate \citep{Neronov}. The maximum VHE emission occurs close to apastron \citep{Veritas, Magic}, suggesting  that absorption plays an important role in the modulation.\\

The outline of the paper is as follows: first we describe the model of attenuation due to pair production. Then
we consider the particular case of LS I +61 303. We assume that the high energy
radiation is emitted from the vicinity of the compact object and that its
emission is isotropic and constant in time. The modulation due to photon-photon interactions is found to be insufficient to account for the VERITAS observations. For this reason, we include  additional interactions of VHE photons with circumstellar material. This model permits to constrain the orbital parameters and the mass of the compact object
as well as the density of ejected material from the companion star.

\section{Interaction with background radiation}
\subsection{Radiative transfer equation}
 The radiative transfer
 equation \citep{chandra} for the intensity $I(s,E)$, where $s$ is the distance
 traveled by a photon of energy $E$ from the emission point is

\begin{equation}
    \frac{dI(s,E)}{ds}=-(1-\cos\xi)\, n(s,\epsilon)\,\sigma(E,\epsilon, \xi)\,I(s,E) +    j(s, E)\,\,; 
    \label{radiative transfer}
\end{equation}

Where $n(s,\epsilon)$ is the spectral density of background photons of energy $\epsilon$
emitted by the main sequence star, $\sigma(E,\epsilon,\xi)$ is the cross
section\footnote{Note that the term $(1-\cos\xi(s'))$ 
corresponds to the relative velocity between the incident and target photons} for the interaction between photons colliding at angle $\xi$, and
$j(E,s)$ is a source term.

\subsection{Neglecting the source term\label{secondaries}}

The source term corresponds to secondary gamma-rays in the
electromagnetic cascade. We can estimate the effect of the source term in the context of a cascade toy model, where the energy of these secondary gamma-rays is degraded
by typically a factor of 4 (after the primary photon first produces an $e^+e^-$ pair). Consequently, the increase in the intensity $I(s, E)$ of gamma rays of energy $E$ 
is increased by $2\tau I(s,4E)+\mathcal{O}(\tau^2)$, where $\tau$ is the probability of interaction (or optical depth as defined in section \ref{solution_section}). Assuming a power spectrum (section \ref{flux_section}) we note that $I(s, 4E)=I(s, E)4^{-\gamma}$. The source term is then approximately given by

\begin{equation}
  j(s,E)\approx 2\times 4^{-\gamma}n(s, \epsilon)\sigma(E,\epsilon, \xi)\,I(s,E).
  \label{source_term}
\end{equation}

In this order of magnitude approximation, we have assumed that the cross section is the same for the energies $E$ and $4E$. In the TeV range, the cross section decreases as $\sim (\epsilon E)^{-1}\ln{\epsilon E}$ \citep{Aharonian}, therefore eq. \ref{source_term} constitutes an upper bound. Now taking $\gamma \sim 2$, we note that the source term is smaller than the attenuation term in eq. \ref{radiative transfer} by a factor of $\lesssim 10 $. Consequently, we neglect the source term (For a detailed numerical calculation, see, e.g. \citet{Sierpowska}).\\

\subsection{Solution of the Radiative transfer equation\label{solution_section}}

With the source term neglected, the solution to the radiative transfer equation is  

\begin{equation}
I(s,E)=I(s_0,E)\,\exp\left\{-\int_{s_0, \epsilon}^{\infty,
    \infty}(1-\cos\xi(s'))\,n(s',\epsilon')\sigma(E, \epsilon',
  s')ds'd\epsilon'\right\}\,\,. 
\label{formal solution}
\end{equation}

Here $s_0$ is the emission point in the vicinity of the compact object (see
figure \ref{orbit}), and $\epsilon$ corresponds to the threshold energy for
pair production,

\begin{equation}
\epsilon=\frac{2m_e^2c^4}{E(1-\cos\xi(s))}.
\end{equation}

The dependence of the
scattering angle $\xi$ in eq. \ref{formal solution} has been changed to a dependence on the path $s$. The problem then reduces to calculating the integral in the exponential of eq. \ref{formal solution}, also known as the optical depth $\tau(s,E)$ \citep{lightman}. In our calculation, we consider the main sequence star as a point source, and in view of the results obtained by \citet{Dubus}, including the angular extension does not change our results significantly. \\

The distribution of background black body photons can be taken as

\begin{equation}
n(r,\epsilon)=n_0(\epsilon)\frac{r_o^2}{r^2},
\label{background}
\end{equation}

where $r_0$ and $n_0$ are the radius of the Be star and the density of
background photons at this radius, i.e.

\begin{equation}
n(r,\epsilon)=\frac{2\pi\,\epsilon^2d\epsilon}{c^3h^3}\left(\frac{r_0}{r}\right)^2\frac{1}{e^{\epsilon/kT}-1}.
\end{equation}

Here, the photon density has already been integrated over the solid angle.

\section{The case of LS I+61 303}
\subsection{Attenuation}

There is debate as to what is the mechanism responsible for high energy emission. However, the aim of this paper is not to model the gamma ray emission but rather to investigate the effects of attenuation. This allows to derive a few characteristics of the main sequence star environment and compact object orbit.\\

\citet{Grundstrom} reported a temperature of 
$T\approx3\times10^{4}\,K$ and radius of $R\approx6.7R_{\odot}$ for the Be star. 
The black body distribution peaks at a few eV, and the threshold energy
for pair production with TeV incident photons is of the order of 1 eV, so that
most of the background photons may contribute to the attenuation,
provided the scattering angle is favorable. The
background photon density is found to be of the order of
$n_{\gamma}\sim10^{12}\,\rm{cm}^{-3}$ at a the radius of the star. The circumstellar
disk has been observed by Waters et al. \citep{waters} and by
Paredes et al. \citep{Paredes}, who estimate the disc ion  density to be
$n_e\sim10^{13}\rm{cm}^{-3}$ at one stellar radius. The cross section for pair production is
of the order of $\sigma_{\gamma\gamma}\approx0.1\sigma_{T}$ at the threshold
energy. The cross section for interaction with hydrogen has a constant value of $\sigma_{\gamma
  H}\approx2\times10^{-2}\sigma_{T}$  above a few hundred MeV \citep{Heitler, Aharonian}. With these cross sections, a first estimate suggests that 
both interactions may result in comparable degrees of attenuation. However, there is a strong angular dependence in the $\gamma\gamma$ interaction, the extreme
case being when the both photons are emitted in the same direction, a configuration in which there is no VHE attenuation. Also the threshold energy is much higher when the incident and target photons are nearly parallel, so fewer
background photons contribute to attenuation. Consequently, $\gamma \gamma$ attenuation may not have strong modulation as a function of the orbital phase 
when compared with the modulation produced by interactions with the circumstellar material.\\

\subsection{Orbital parameters of $LS\,I\,+61^\circ 303$}
\begin{figure}[h!]
  \begin{center}
    \includegraphics[scale=0.5]{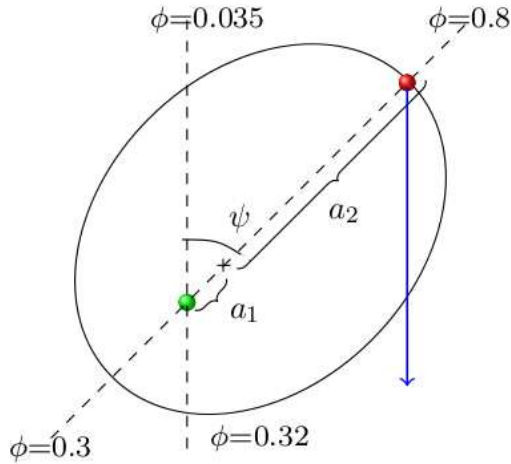}
  \end{center}
  \caption{\label{orbit}Sketch of the orbital parameters of $LS\,I\,+61^\circ 303$ for the case of $i=90^{\circ}$. The
    arrow points to the observer. Also shown are the values of the phase at periastron, apastron, and superior and inferior conjuction.} 
\end{figure}

The orbital parameters of $LS\,I\,+61^\circ 303$, as sketched in figure \ref{orbit}, are still subject of research 
\citep{Aragona, Grundstrom, Casares}. The quantities of interest 
are: the period $P$, the angle between the major axis of the
ellipse and the line of sight $\psi$, the projected
semi-major axis ($a_1\sin i$), corresponding to the ellipse of the Be star\footnote{The projected semi-major axis of the ellipse described by the compact object is typically labeled as $a_2\sin i$.}, the
eccentricity $\eta$, the phase at periastron $\phi_0$, and the mass function $f(M_1,M_2)$,
which depends on the period and the radial velocity and relates the masses of
both objects and the inclination angle $i$. The most recent orbital solution
has been 
obtained by \citet{Aragona}, where $P=26.4960\,d$,
$\psi=40.5\pm5.7^{\circ}$, $a_1\sin i=8.64\pm0.52\,R_{\odot}$,
$\eta=0.54\pm0.03$, $\phi_0=0.275$ and
$f(M_1,M_2)=0.0124\pm0.0022\,M_{\odot}$. Assuming the attenuation is responsible for the periodic modulation of the TeV emission, our results constrain  $i$ and thereby the mass of the compact object.

\section{The integrated flux\label{flux_section}}

Following observations of $LSI+61^\circ 303$ from 09/2006 to 02/2008, the VERITAS collaboration reported power law spectrum ($\frac{d\Phi}{dE}=\Phi_0\left(\frac{E}{TeV}\right)^{-\gamma}$) with a spectral index
of $\gamma=2.4\pm0.2_{stat}\pm0.2_{syst}$ at energies above $\sim0.5\,\rm{TeV}$,  and between phases $\phi=0.6$ and
$\phi=0.8$ \citep{Veritas}. Observations at lower energies made by Fermi between 08/2008 and 03/2009, indicate that the
spectral index does not change significantly as a function of the orbital
phase \citep{Fermi}. Therefore, we assume a constant
intrinsic\footnote{By intrinsic we mean non attenuated by pair production.}
spectrum as a function of the phase at TeV energies. The integrated flux is then

\begin{equation}
F(\phi)=\int_{E_0}^{\infty}\frac{d^3N}{dEdtdA}
I(E,\phi)dE=F_0\int_{E_0}^{\infty}\left(\frac{E}{E_0}\right)^{-\gamma}
I(E,\phi)dE,
\label{integrated flux eq}
\end{equation}

where $E_0$ depends on the 
detection threshold energy of the detector, and $F_0$ is a normalization factor that is taken as 
a free parameter. 

\subsection{Light curve assuming only $\gamma\gamma$ interactions\label{gamma-gamma}}

\begin{figure}[!h]
  \begin{center}
    \rotatebox{-90}{\includegraphics[scale=0.5]{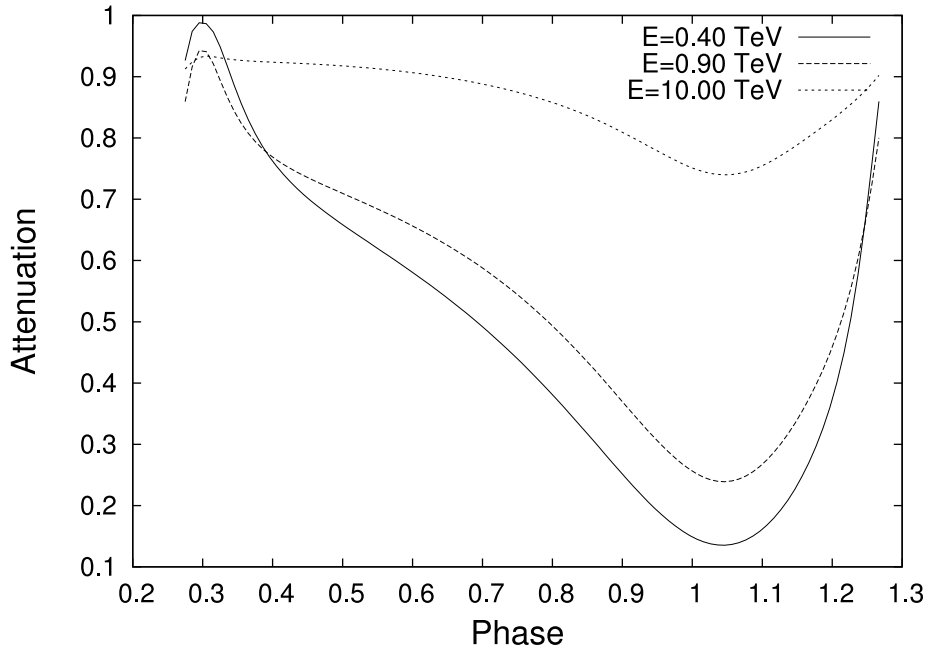}}
    \caption{\label{att_gamma_gamma} Attenuation $e^{-\tau_{\gamma\gamma}}$ as a function of the orbital phase for 
      different primary photon energies ($\gamma\gamma$ interactions only). 
      A mass of $1.5M_{\odot}$, corresponding to $i=64^{\circ}$, was assumed for the compact object.}
  \end{center}
\end{figure}

Figure \ref{att_gamma_gamma} shows the attenuation as a function of the orbital phase
for several different energies for the case of the compact object having the
canonical neutron star mass ($i\approx 64^{\circ}$ or $M\approx 1.5M_\odot$). In figure \ref{att_gamma_gamma} we
essentially reproduce one of the results obtained by Dubus, except that the
orbital parameters used are the newer set obtained by \citet{Aragona}.
When only interactions with the background black body photons are taken into account, and the orbital plane is closer to being seen edge-on,
the optical depth  approaches a minimum when the compact
object is close to the main sequence star. This is especially the case for very 
high inclination angles, corresponding to the mass of the compact object being small, and
close to the Chandrasekhar mass. This behavior can be understood from
the angular dependence of the threshold energy in addition to the  
relative velocity of the incident and target photons approaching a minimum.  Also, at
high energies, the cross section for pair creation decreases as the inverse square of
the center of mass energy, decreasing the optical depth even more. That is,
even though the total density of background photons  
increases (as $1/r^2$) when the compact object approaches the Be star, a combination
of the previously mentioned factors dominates as can seen in figure
\ref{att_gamma_gamma}.\\
 
\begin{figure}[!h]
  \begin{center}
    \rotatebox{-90}{\includegraphics[scale=0.5]{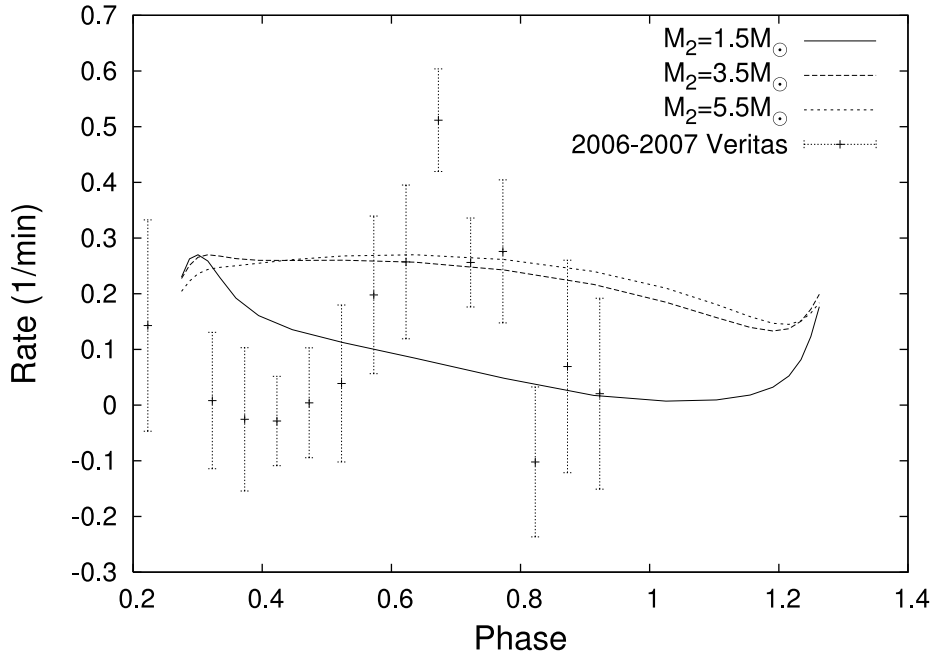}}
    \vspace{1cm}
    \caption{\label{light_curve_gamma_gamma}Normalized light-curve for $\gamma\gamma$ interactions only. 
      Each curve corresponds to a different mass of the compact object and inclination angle.}
  \end{center}
\end{figure}

Figure \ref{light_curve_gamma_gamma} shows the normalized integrated flux
assuming different inclination angles and corresponding compact object masses. The VERITAS data shown in figure
\ref{light_curve_gamma_gamma} \citep{Veritas2} were binned to show a single light-curve 
as opposed to monthly data. If we assume that the emission comes from the vicinity  
of the compact object, and is isotropic, and constant as a function of 
the orbital phase, then these  results lead us to conclude that there must be an additional  
attenuation mechanism at play.\\

\subsection{Light curve including $\gamma\gamma$ and $\gamma H$ interactions\label{gamma-h}}

The detailed structure of the circumstellar material surrounding a Be star in
the presence of a compact companion has been studied by \citet{waters}, \citet{marti} and \citet{reig} among others. It is thought to consist of a main 
equatorial disk-like component and a polar wind. Typically, the parameters that describe the decretion disk include:  
the mass loss rate, the wind termination velocity, the half opening angle of the disk, and  
radius of the disk. The polar wind is radiatively driven and may have a corresponding mass loss rate comparable to the equatorial wind (within one order of magnitude \citep{waters}). When comparing the quality of the data shown in figure \ref{light_curve_gamma_gamma}, and the complexity 
of the models that describe the circumstellar material, it appears that, when only TeV data is used, only an order of magnitude estimate of
the material density and its extension in the system can be achieved. With this in mind, 
we rather assume a simple isotropic distribution of material that decreases as
a power $q$ of the 
distance from the Be star ($n=n_H(r_0/r)^q$). We start by setting $q=2$ and then consider different values for comparison. Parameters found from existing models are 
 taken into consideration for our approximation.\\

\begin{figure}
\begin{center}
  \rotatebox{-90}{\includegraphics[scale=0.5]{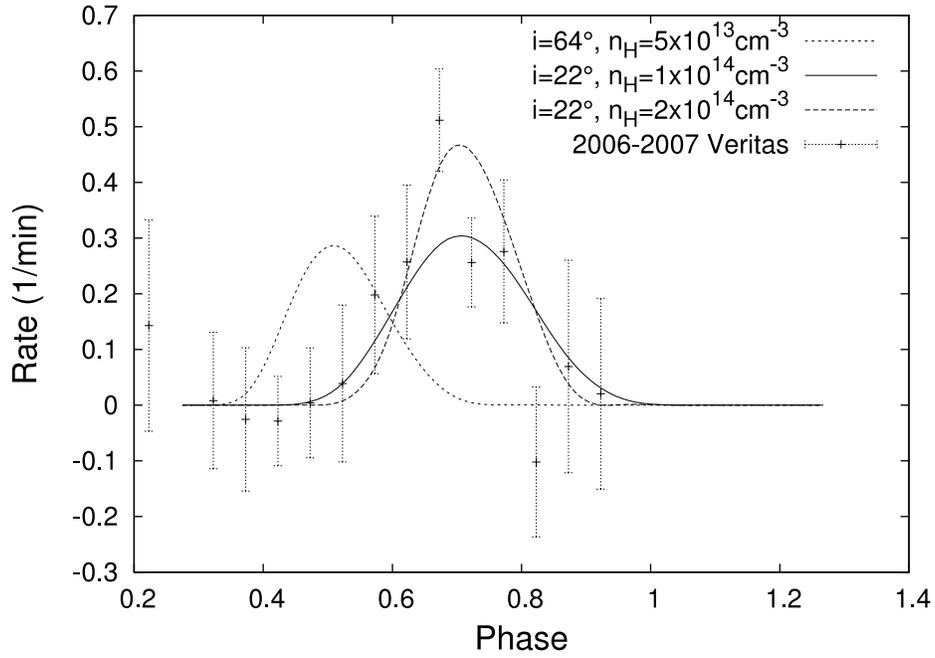}} 
  \vspace{1cm}
  \caption{\label{light_curve_gamma_h} Light curves for different inclination angles $i$ and hydrogen densities $n_H$ when an isotropic
    distribution of hydrogen is included. The mass of
    the compact object is $4.0M_{\odot}$ for the two curves whose peak is
    at phase $\sim 0.7$, and $1.5M_{\odot}$ for the curve whose peak is closer
    to phase $\sim 0.4$. The concentration of hydrogen at $r\approx100R_{\odot}$ for
    each curve is labeled in the top right-hand corner of the figure. In the two right-most curves, the intrinsic luminosity (normalization factor) 
  is chosen so as to maximize agreement with data, and it increases with the hydrogen density.}
  \end{center}
\end{figure}

 For the case of a constant cross
 section and a $1/r^2$ distribution of hydrogen, the optical depth can
 actually be found analytically (see appendix A). The resulting light curves are shown in figure \ref{light_curve_gamma_h} 
 for different values of the inclination angle. For low inclination angles (high mass), VHE photons emitted from apastron go through less circumstellar hydrogen than those emitted from periastron. This results in minimum attenuation near apastron, as can be seen in figure \ref{light_curve_gamma_h}. From this figure it is clear that the emission
 peak corresponding to a canonical $1.5M_{\odot}$ neutron star 
 is only marginally supported by observations. \\

We take the inclination angle, characteristic density, and normalization factor as free
 parameters. We find the inclination angle to be $i< 28^\circ$ ($M_2>3M_\odot$)
 at the $89\%$ confidence level (CL) in the
context of this model, and $i<34^{\circ}$ ($M_2>2.5M_{\odot}$) at the $99\%$
CL. These limits are not in good agreement with the
neutron star scenario generally favored for the broad-band spectrum it
implies\footnote{See Zdziarski et al. \citep{Neronov} for more details}. However, our results are   
still consistent with other observational constraints ($10^{\circ}<i<60^{\circ}$) \citep{Casares} obtained from optical spectroscopy. As for
the circumstellar material, if we assume the characteristic extension to be
$r_0\approx100R_\odot$, consistent with more sophisticated models
\citep{Torres}, then the density of hydrogen in the disk is found to be $2.0\times 10^{13}\rm{cm}^{-3}\leq n_H\leq 1.9\times 10^{15} \rm{cm}^{-3}$ at the $99\%$ CL and $n_H=\left(2.7\pm ^{11.3}_{2.1}\right)\times 10^{14}\rm{cm}^{-3}$ at the $68\%$ CL.\\

By integrating the volume density along the line of sight to the compact object at apastron, we find a column density of $1.9\times10^{26}\rm{cm}^{-2}\leq N_H \leq 1.8\times 10^{28}\rm{cm}^{-2}$ at the $99\%$ CL, which is much higher than results found elsewhere in the literature \citep{waters, marti, esposito}. In particular, when we use the column density found by X-ray observations $N_H=(5.7\pm 0.3)\times 10^{21}\rm{cm}^{-2}$ \cite{esposito}, we find a reduced $\tilde{\chi}^2$ of $3.06$ (11 degrees of freedom), corresponding to a $\chi^2$ probability $P(\tilde{\chi}^2\geq 3.06)=0.04\%$. A rough estimate suggests that by including $\sim 10\%$ of helium, the column density would be reduced by a factor of $\sim 2$, which is not sufficient to achieve compatibility with X-ray results. \\

Density profiles in Be stars typically have radial dependences of $1/r^q$, where $2.3<q<3.3$ \citep{Lamers}, depending on the opening angle of the disk. Therefore we expect our constraint on the density to constitute a lower bound\footnote{This is assuming that the disk and orbit lie in the same plane.}. We perform our calculation with $q=3$ and note that our results do not change considerably.\\ 

The hydrogen density also corresponds to a mass loss rate of $\dot{M_1}\approx10^{-7}\Omega \frac{V_{\mathrm{wind}}}{\mathrm{1km\,s^{-1}}} M_\odot \rm{yr}^{-1}$, where $\Omega$ is the solid angle. Typically accepted values for the mass loss rate are in the range of $\sim 10^{-7}M_{\odot}\rm{yr}^{-1}$ to $10^{-8}M_{\odot}\rm{yr}^{-1}$, 
 as have been reported by \citet{snow} and \citet{waters}
 among others. A first glance at our result for the mass loss implies that it
 does not agree with the observations, i.e. setting $\Omega=4\pi$ and
 $V_{wind}\sim 100\rm{km\,s}^{-1}$ \citep{waters}. However, if we relax the assumption of
 an isotropic distribution of hydrogen, our result implies that small solid
 angles are favored as well as small velocities for the stellar wind. Small
 solid angles are consistent with the thin disk scenario that is most commonly
 accepted. Small velocities of the order of a few $\rm{km\,s^{-1}}$ are however not
 consistent with what is found elsewhere in the literature,
 e.g. \citep{waters}, and the wind indeed has higher velocities, this would imply that the system may have been observed while in a state of high mass loss rate. \\

\section{Discussion}

Since, in the TeV range, the interaction with matter is approximately independent of the energy, 
and since, as figure \ref{light_curve_gamma_gamma} shows, $\gamma\gamma$ interactions are insufficient 
to account for the orbital modulation, then the intrinsic non-attenuated differential spectrum
 is essentially the same as the observed spectrum (a power law of
spectral index $-2.4$). However, the intrinsic TeV luminosity is several orders
of magnitude higher than the measured luminosity. Taking the distance to the
source to be approximately $1.8\,\rm{kpc}$ \citep{Steele}, we find the intrinsic luminosity to be $L\approx5\times10^{37}\rm{erg\,s^{-1}}$ when the hydrogen density is of the order of $\sim5\times 10^{13}\rm{cm}^{-3}$. This intrinsic luminosity is comparable to that suggested by  \citet{Bottcher} for \emph{LS 5039}, the only other known TeV binary thought to contain a black hole.\\

It is interesting to compare this intrinsic luminosity to the Eddington
luminosity\footnote{At the energies considered here, the cross section for
  inverse Compton is $\sim 0.1\sigma_T$}
$L_{Edd}\approx1.3\times10^{39}(M_2/M_\odot)\rm{erg\,s^{-1}}$, which is comparable to $L$, and implies that 
radiation may be beamed in our direction. It is also
interesting to calculate the accretion rate that would be needed in order to
obtain the intrinsic luminosity: By taking $L\approx GM_2\dot{M_2}/R$, where $R$
is of the order of the Schwarzschild radius ($2GM_2/c^2$), we find
$\dot{M_2}\approx 2\times 10^{-8}M_\odot \rm{yr}^{-1}$. This rate is  comparable with the observed mass loss rate of $\sim 10^{-8}M_\odot \rm{yr}^{-1}$. The
fact that the accretion rate is comparable to the measured mass loss rate, suggests that
the flow of matter can be quite complicated, e.g. an increase in the accretion
rate would strip most of the circumstellar mass, leading to time variability. This 
may explain
the fact that no VHE detections have been reported since 2008.\\

Still assuming the intrinsic luminosity to be constant in time, we can estimate the amount of hydrogen needed 
 to attenuate the source to below the detectability threshold. We find that the
density must increase from $\sim5\times10^{13}\rm{cm}^{-3}$ to
$\sim5\times10^{14}\rm{cm}^{-3}$ at the characteristic distance of $100R_{\odot}$. This amount of hydrogen in turn leads
to much higher mass loss rates than those observed, and it may also imply a stronger activity of the source.\\

It is worth mentioning that the attenuation model is not the only possible way to account for 
 the modulation. For example, there is also the possibility of the emission being anisotropic, and the modulation
 resulting from a geometrical effect. This possibility is described in detail by \citet{Neronov}, were a shocked 
pulsar wind with a large Lorentz factor is thought to be the cause of emission.

\section{Conclusions}

For the case of $LS\,I\,+61^\circ303$,
we find that attenuation due to $\gamma\gamma$ interactions with the
background radiation does not account for the observed high energy
flux modulation as a function of the orbital phase, namely a narrow peak
near apastron. This effect leads us to investigate some
properties of the ejected material from the Be star, and the inclination angle of the orbit. We find the
angle of the orbit to be $i<34^\circ$ ($M>2.5M_\odot$) at the $99\%$ confidence level, 
suggesting that the compact object is a black hole rather than a neutron star. We
also find the density of hydrogen in the disk to be 
$2\times 10^{13}\rm{cm}^{-3}\leq n_H\leq 2\times 10^{15} \rm{cm}^{-3}$ at the $99\%$ CL (at $100R_\odot$), which accounts for most of the
observed gamma ray absorption.  
If the compact object is indeed a black hole as our analysis
suggests, then the gamma ray emission is likely to be powered by accretion \citep{bosch, Neronov}.
Also, a black hole scenario might be even more complicated due to
the possibility of VHE emission originating from termination of jets, therefore we
cannot exclude the possibility of the modulation being due to geometrical effects.  Current VHE data does not allow to constrain the system much more
than what we have already done, and the fact that VHE detections have not been
reported since the VERITAS \citep{Veritas} and MAGIC \citep{Magic} detections
where made, makes the problem even more puzzling. A possible explanation might
originate from a complex matter flow. This is suggested by the fact that the accretion rate needed to explain an intrinsic
non-attenuated luminosity, is comparable to the measured mass loss rate of the Be star.\\

An inconsistency arises when comparing our results with those derived from X-ray observations. We find the column density to be $1.9\times10^{26}\rm{cm}^{-2}\leq N_H \leq 1.8\times 10^{28}\rm{cm}^{-2}$ ($99\%$ CL), which is only compatible with X-ray results at the 0.04\% confidence level. Such an incompatibility may imply that pair production in the stellar wind is not the cause of the modulation. Consequently, our estimates on the mass and column density may not be valid. An alternative explanation by \citet{Neronov} suggests that the modulation is due to a geometrical effect. Here a shocked pulsar wind is thought to flow along a cone with a large Lorentz factor, producing beamed radiation which can be seen when the cone passes through the line of sight.

\section*{Appendix A: Optical depth for constant cross section and $1/r^2$ density
  distribution}

\begin{figure}
  \begin{center}
    $
    \includegraphics[scale=1.4]{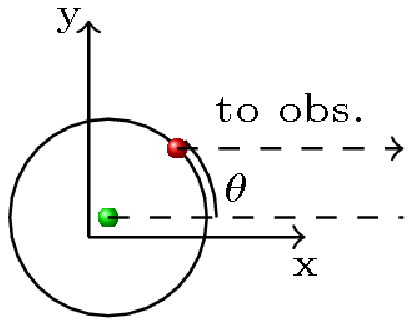}
    \includegraphics[scale=0.3]{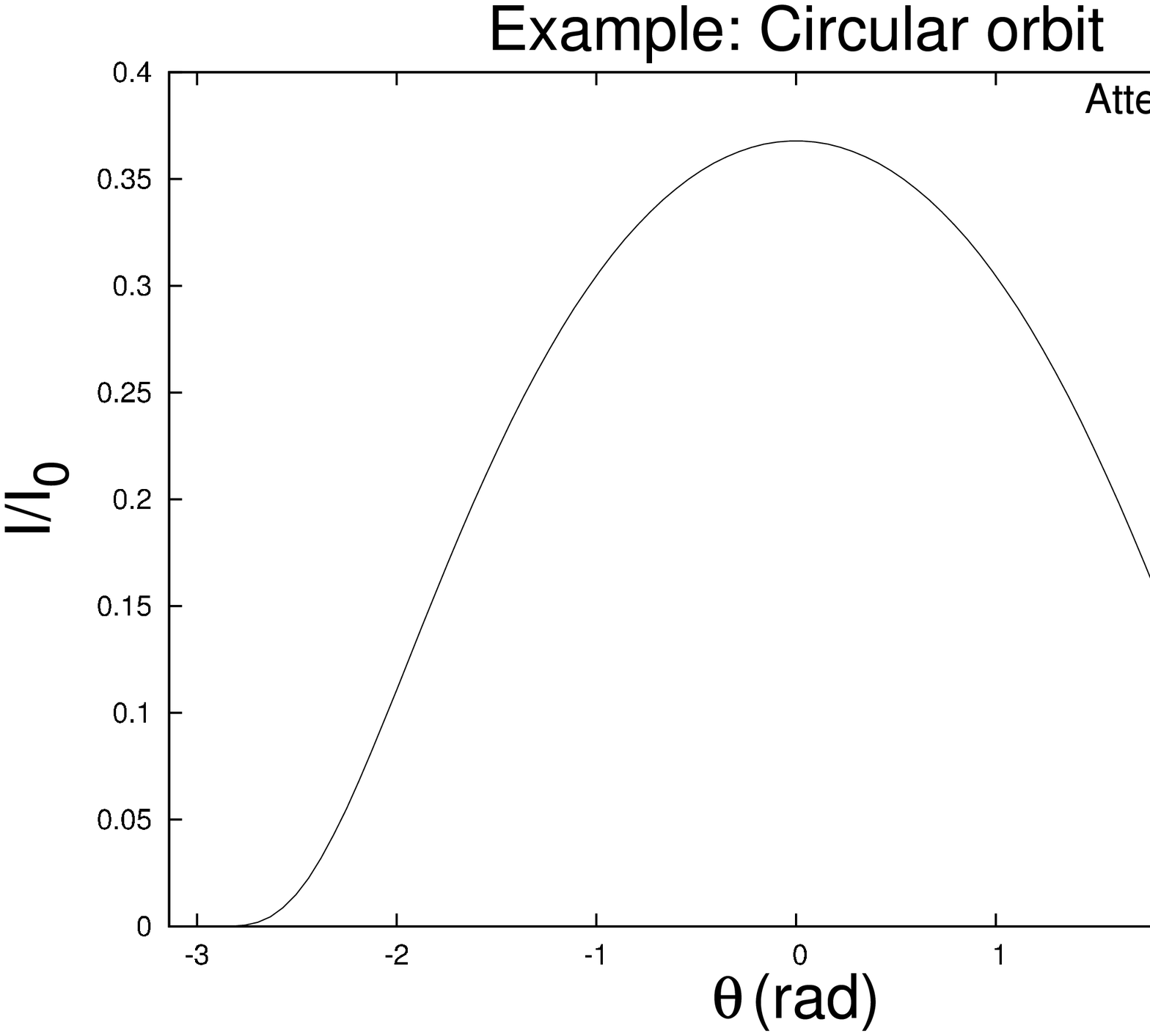}
    $
\end{center} 
 \caption{\label{circle}  Coordinate system used for calculating the optical
    depth  (eq. \ref{first approximation}).}
\end{figure}

Using a $1/r^2$ distribution of hydrogen, the cross section $\sigma_H$
accounting for interactions between VHE photons and hydrogen, and the system
of coordinates shown in figure \ref{circle} (corresponding to an orbital plane
seen edge on) , we can calculate the integral for the optical depth to be

\begin{eqnarray}
  \int_{x_i}^{\infty}\frac{n_Hr_0^2\sigma_H}{x^2+y_i^2+z_i^2}dx 
  &=&\left[\frac{n_Hr_0^2\sigma_H}{\sqrt{y_i^2+z_i^2}}\tan^{-1}\left(\frac{x}{\sqrt{y_i^2+z_i^2}}\right)\right]^{\infty}_{x_i}\\
  &=& \frac{n_Hr_0^2\sigma_H}{\sqrt{y_i^2+z_i^2}}\left(\frac{\pi}{2}-\tan^{-1}\left(\frac{x_i}{\sqrt{y_i^2+z_i^2}}\right) \right),
  \label{first approximation}
\end{eqnarray}

where $x_i$, $y_i$, and $z_i$ are functions of the orbital angle $\theta$, and $r_0$
is the characteristic radius of the hydrogen disk. For the case of a circular orbit as seen edge on
(figure \ref{circle}), we can easily see the limiting behavior of the
intensity as a function of the orbital angle. That is, expanding around $\theta\sim0$ reveals that the attenuation around this region behaves like a Gaussian.

\begin{equation}
For\,\,\theta\sim0:\,\,I(r_i, \theta)= I_0(\theta, r_i)e^{-\frac{n_Hr_0^2\sigma_H}{r_i}(1+\theta^2)}.
\end{equation}

Similarly, expanding around $\theta\sim\pi/2$ reveals that the attenuation behaves like a decreasing exponential

\begin{equation}
For\,\,\theta\sim\pi/2:\,\,I(r_i, \theta)=I_0(\theta, r_i)e^{-\frac{n_Hr_0^2\sigma_H}{r_i}\theta}
\end{equation}

For a more complicated geometry of $LS\,I\,+61^\circ 303$, it is now just a matter of inserting the appropriate expressions for $x_i(\theta)$, $y_i(\theta)$ and $z_i(\theta)$.

\end{document}